\documentclass[%
  reprint, twocolumn, %HERE (uncomment here for reprint)
  showpacs,
  superscriptaddress,
  amsmath,amssymb,
  aps,
  prb,
]{revtex4}

\usepackage{textcomp,color}
\usepackage{graphicx}% Include figure files
\usepackage{bm}% bold math
\begin{document}

%\preprint{}

\title{%
  Crossover Behavior from Decoupled Criticality}


\author{Y. Kamiya}
%\thanks{}
\author{N. Kawashima}
%\thanks{}
\affiliation{%
  Institute for Solid State Physics, University of Tokyo, 
  Kashiwa, Chiba 227-8581, Japan
}
\author{C. D. Batista}
\affiliation{%
  Theoretical Division, Los Alamos National Laboratory, 
  Los Alamos, New Mexico 87545, USA
}


\date{\today}

\begin{abstract}
  We study the thermodynamic phase transition of a spin Hamiltonian comprising
  two 3D magnetic sublattices. Each sublattice contains XY spins coupled by
  the usual bilinear exchange, while spins in different sublattices only
  interact via biquadratic exchange. This Hamiltonian is an effective model
  for XY magnets on certain frustrated lattices such as body centered
  tetragonal. By performing a cluster Monte Carlo simulation, we investigate
  the crossover from the 3D-XY fixed point (decoupled sublattices) and find a
  systematic flow toward a first-order transition without a separatrix or a
  new fixed point. This strongly suggests that the correct asymptotic behavior
  is a first-order transition.
\end{abstract}

\pacs{75.30.Kz, 05.70.Jk, 05.50.+q, 75.40.Mg}
%\keywords{}

\maketitle


\section{Introduction}
Geometric frustration can play a decisive role in the behavior of magnetic
systems. The combination of frustrated geometries with strong quantum
fluctuations can lead to new quantum states of matter 
\cite{kageyama1999exact, kodama2002magnetic, sebastian2008fractalization}. 
It has been shown recently that novel charge effects in Mott insulators, such
as spin-driven electronic charge density waves, or orbital currents, only take
place in geometrically frustrated lattices \cite{bulaevskii2008electronic}. 
Geometric frustration can also reduce the effective dimensionality of certain
quantum critical points \cite{sebastian2006dimensional}. Finally, it has been
known for years that the presence of geometric frustration can change the
nature of certain thermodynamic phase transitions. However, it has been also
recognized that the nature of the new transition can be very elusive for the
standard renormalization group treatments \cite{delamotte2008fixed} 
and may require very sophisticated numerical approaches 
\cite{hukushima2000random, itakura2003monte}.

Several quantum magnets comprise two sublattices of magnetic ions coupled by a
geometrically frustrated exchange \cite{sebastian2006dimensional}. This is for
instance the case of a Heisenberg antiferromagnet on a body centered
tetragonal (BCT) lattice \cite{kamiya2009finite}, or a square lattice with
nearest- and next-nearest-neighbor exchange interactions
\cite{henley1989ordering, chandra1990ising}. We are interested in the regime
of inter-sublattice coupling smaller than the intra-sublattice exchange. We
will also assume that there is a uniaxial easy-plane anisotropy that reduces
the Hamiltonian symmetry from O(3) to O(2). The frustrated nature of the
inter-sublattice exchange precludes a bilinear coupling between the order
parameters of the two sublattices. The Hamiltonian symmetry only allows for an
effective {\it biquadratic} coupling. Consequently, if $\bm{\mathrm{m}}_{A}$
and $\bm{\mathrm{m}}_{B}$ are the XY magnetizations at wave-vector $\bm{k}_{0}
= (\pi, \pi, 0)$ of the sublattices (which in the BCT lattice case are the
even- and odd-numbered layers\cite{kamiya2009finite}), 
the parallel or anti-parallel orientations of 
$\bm{\mathrm{m}}_{A}$ and $\bm{\mathrm{m}}_{B}$ correspond to different ground
states. The Z$_2$ symmetry is broken by selecting one of these two states 
\cite{batista2007geometric, rosch2007quantum, schmalian2008emergent, kamiya2009finite}. 
The O(2)$\times$Z$_2$ symmetry breaking also appears in the XY model on a
triangular lattice \cite{hasenbusch2005multicritical}. 
In this case the  $Z_2$ broken symmetry corresponds to the two possible vector chiral orderings.

We want to explore the nature of the thermodynamic phase transition 
associated with the O(2)$\times$Z$_2$ symmetry breaking that takes place 
in several frustrated magnets. For this purpose, we will consider 
classical magnetic moments because
the quantum character of the spins does not affect the nature of the
thermodynamic transition.
%In the following, assuming that the primary contribution to the
%statistical weight near the finite temperature transition is given
%by configurations with small fluctuation in the imaginary-time 
%direction (in the coherent-state representation), we consider 
%classical spins. 
%As argued in Ref. \cite{kamiya2009finite}, 
%the effective interlayer interactions generated via thermal
%fluctuations are ferro-quadrupolar or biquadratic  
%between nearest-neighbor (NN) layers and ferromagnetic 
%between next-nearest-neighbor layers. A bilinear coupling
%between NN layers is prohibited by the Z$_2$ lattice-symmetry. 
In Ref.~\onlinecite{kamiya2009finite}, we used two different
approaches to understand the effect of the additional Z$_2$ symmetry
breaking and compared their results. The first approach was a Monte Carlo (MC) simulation of the
classical spin model on the BCT lattice. The second approach was a scaling
analysis of the Landau-Ginzburg-Wilson (LGW) model that preserves the
symmetries of the lattice Hamiltonian. A single transition with exponents
close to those of the 3D XY model was obtained from a finite-size scaling
(FSS) analysis of the MC data \cite{kamiya2009finite}. On the other hand, the
scaling analysis of the LGW model,
\begin{multline}
  H_{\text{LGW}} = \int d^{d}x\biggl[\,
    \sum_{a = A, B}
    \left(
    \frac{1}{2}
    \left|\nabla{\phi}_{a}\right|^2
    +t\lvert{\phi}_{a}\rvert^2
    +u\lvert{\phi}_{a}\rvert^4
    \right)
    \\ %HERE (uncomment here for reprint)
    +\lambda
    \left({\phi}_{A}\cdot{\phi}_{B}\right)^2
    +g
    \lvert{\phi}_{A}\rvert^2 \lvert{\phi}_{B}\rvert^2
    \biggr],
  \label{eq:lgw-hamiltonian}
\end{multline}
indicated that $\lambda$ is a relevant perturbation for the 3D XY decoupled
fixed point (DFP) located on the $u$ axis ($u \ne 0$, $\lambda = g = 0$)
\cite{kamiya2009finite}. 
Here, 
$
\phi_{a} = \left(\phi_{a}^{x}, \phi_{a}^{y}\right)
$ 
($a = A, B$) is a two-component field representing antiferromagnetic moments
in even- ($a = A$) or odd- ($a = B$) numbered layers, and $\lambda$ is the
biquadratic coupling between them. These results look contradicting at a first
glance: although the numerical observations can be explained in a consistent
way by the DFP, this fixed point is nevertheless \textit{unstable} along the
$\lambda$-direction. More specifically, near the DFP, $\lambda$ transforms as 
$
\lambda' = b^{y_{\lambda}}\lambda,
$
where
$
y_{\lambda} = 0.526(8)
$
and $b$ is a rescaling factor \cite{kamiya2009finite}.

The scaling argument implies that there will be a crossover behavior from the
DFP, provided
$
\left|\lambda\right| L^{y_{\lambda}}\gtrsim 1
$
with $L$ being the system-size \cite{kamiya2009finite}. However,
$\left|\lambda\right|$ can be quite small for the original frustrated spin
system because it is an effective interaction that arises from second-order
perturbation with respect to the ratio between the inter- and the intra-layer
bilinear exchange couplings \cite{kamiya2009finite}. In addition, we could not
obtain data for sufficiently large $L$ in our previous calculation in Ref. \onlinecite{kamiya2009finite} because we
simulated the original Hamiltonian on the {\it frustrated} lattice. 
Thus, the nature of the crossover was left as an open
problem.

\section{Model and Method}
\subsection{Model}
In this paper, we explore the expected 
crossover by studying an XY spin model on a cubic lattice that is 
more directly related to the LGW effective model than to the original Hamiltonian on
the BCT lattice. The relevant coupling $\lambda$ is explicitly taken into 
account by considering the Hamiltonian model:
\begin{equation}
  H = -J\sum_{\left\langle i,j\right\rangle, a=A,B}
  \bm{\mathrm{S}}_{a, i}\cdot\bm{\mathrm{S}}_{a, j}
  +\lambda J\sum_{i}
  \left(\bm{\mathrm{S}}_{A, i}\cdot\bm{\mathrm{S}}_{B, i}\right)^{2}
  \label{eq:lattice-hamiltonian}
\end{equation}
with $J > 0$. $\bm{\mathrm{S}}_{a, i}$
($a = A, B$) is a classical XY spin at site $i$ on the cubic 
lattice and $\left\langle i,j\right\rangle$ is a pair of 
nearest-neighbor sites. The coefficient $\lambda$ characterizes the
amplitude of the biquadratic coupling that is expected to drive the
system away from the DFP. We consider the case $\lambda < 0$, which 
is experimentally relevant \cite{kamiya2009finite}. No term corresponding to the $g$-term 
in $H_{\text{LGW}}$ is explicitly included in $H$, because it is automatically
generated when the short wavelength modes are integrated out (renormalization process).

In the ground state, 
both $A$ and $B$ spins are ferromagnetically ordered and the O(2) 
symmetry is broken. In addition, their relative phase is locked so
that  $\bm{\mathrm{S}}_{A, i}\cdot\bm{\mathrm{S}}_{B, i} = \pm 1$,
which causes Z$_{2}$ symmetry breaking. The order parameters
associated with these two kinds of symmetry breaking are
$\bm{\mathrm{m}} = \bm{\mathrm{m}}_{A}$ with
$\bm{\mathrm{m}}_{A} = L^{-d}\sum_{i}\bm{\mathrm{S}}_{a, i}$ and
$\sigma = L^{-d}\sum_{i}\sigma_{i}$ with 
$\sigma_{i} = \bm{\mathrm{S}}_{A, i}\cdot\bm{\mathrm{S}}_{B, i}$,
respectively. We introduce the correlation functions
$
G_{ij}^{m}
= \left\langle\bm{\mathrm{S}}_{A, i}\cdot\bm{\mathrm{S}}_{A, j}\right\rangle
$
and
$
G_{ij}^{\sigma}
= \left\langle\sigma_{i}\sigma_{j}\right\rangle
$.
(For the definition of $\bm{\mathrm{m}}$ and $G_{ij}^{m}$, we can use either $A$ or $B$
spins without loss of generality.)
%Based on our numerical evidence, 
%$
%\left\langle|\bm{\mathrm{m}}_{A}|^2\right\rangle
%= \left\langle|\bm{\mathrm{m}}_{B}|^2\right\rangle
%$,
%$
%\left\langle|\bm{\mathrm{m}}_{A}|^4\right\rangle
%= \left\langle|\bm{\mathrm{m}}_{B}|^4\right\rangle
%$,
%$
%\left\langle\bm{\mathrm{S}}_{A, i}\cdot\bm{\mathrm{S}}_{A, j}\right\rangle
%= \left\langle\bm{\mathrm{S}}_{B, i}\cdot\bm{\mathrm{S}}_{B, j}\right\rangle
%$, 
%we assume that the symmetry between the two sublattices remains unbroken.

For very small $\left|\lambda\right|$ 
($\lambda = -0.05$, $L \le 64$), we observe an apparently 
continuous transition with exponents of the DFP, which is naturally
interpreted as the same behavior as in the previous MC simulation in
Ref.~\onlinecite{kamiya2009finite}.
However, a more careful FSS analysis reveals the expected 
crossover. We present a numerically obtained renormalization-group 
flow diagram of several scaling parameters 
that should be scale-invariant at the second-order transitions
\cite{hukushima2000random, itakura2003monte}. 
We find that the flow evolves systematically from the DFP without a sign of a
stable fixed point or a separatrix, toward the region where
the transition is discontinuous. Based on this 
observation and the lack of a stable fixed point in the $\epsilon$-expansion 
($\epsilon = 4 - d$) around the DFP \cite{aharony1975critical}, we 
propose that the correct asymptotic behavior is a first-order transition
for any (negative) finite value of $\lambda$.

\subsection{Method}
The absence of explicit frustration is the main computational advantage of $H$
relative to the original model studied in Ref.~\onlinecite{kamiya2009finite}. 
This enables us to develop an 
efficient cluster MC algorithm based on a minor modification of the embedding method proposed by 
Wolff \cite{wolff1989collective}. In every update cycle, we choose a unit vector $\bm{\mathrm{n}}$ at 
random. The vector $\bm{\mathrm{n}}$ defines the Z$_2$ transformations
$
\bar{\bm{\mathrm{S}}}_{A} 
= \bm{\mathrm{S}}_{A} - 2\left(\bm{\mathrm{S}}_{A}\cdot\bm{\mathrm{n}}\right)\bm{\mathrm{n}}
$
and 
$
\bar{\bm{\mathrm{S}}}_{B}
= -\bm{\mathrm{S}}_{B} + 2\left(\bm{\mathrm{S}}_{B}\cdot\bm{\mathrm{n}}\right)\bm{\mathrm{n}}
$.
(The difference by a factor of $-1$ serves to enhance the 
relaxation of the $\sigma$ modes as compared to applying the same 
mirror-image transformation to the $A$ and $B$ spins.) Then, we 
choose a spin $\bm{\mathrm{S}}_{a, i}$ and identify a cluster
$C = \{\bm{\mathrm{S}}_{a, i}, {\bm{\mathrm{S}}}_{b, j}, {\bm{\mathrm{S}}}_{c, k}, \dots\}$ 
that can be reached from $\bm{\mathrm{S}}_{a, i}$ via 
probabilistically activated links. The probability to activate a link
depends on the interaction on the link:
$ 
P_1(\bm{\mathrm{S}}, \bm{\mathrm{S}}')
= 1 - \min\left\{
1, \exp\left[\beta J\left(
  \bar{\bm{\mathrm{S}}} - \bm{\mathrm{S}}
  \right)\cdot\bm{\mathrm{S}}'
  \right]
\right\}
$
for links with the bilinear exchange and
$
P_2(\bm{\mathrm{S}}, \bm{\mathrm{S}}')
= 1 - \min\left(
1, \exp\left\{
\left|\lambda\right|\beta J
\left[
    \left(\bar{\bm{\mathrm{S}}}\cdot \bm{\mathrm{S}}'\right)^2
    - \left(\bm{\mathrm{S}}\cdot \bm{\mathrm{S}}'\right)^2
    \right]
\right\}
\right)
$
for links with the biquadratic coupling. After a cluster is 
identified, we flip it, namely apply the Z$_2$ transformation 
on every spin included in $C$. It can be easily checked that the 
algorithm satisfies both the detailed-balance and ergodicity 
conditions.

\section{Results}
\subsection{Conventional scaling analysis}
\begin{figure}
  \centering
  \includegraphics[width=8.6cm]{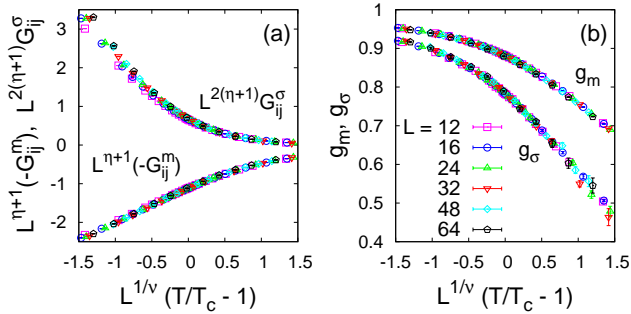}   %HERE (uncomment here for reprint)
  \caption{
    (Color online)
    ``Pseudo-scaling'' behavior observed for $\lambda = -0.05$ 
    ($\left|\lambda\right|L^{y_{\lambda}} \approx 0.45$ for $L = 64$) of
    (a) correlation functions at a distance
    $r_{ij,x} = r_{ij,y} = r_{ij,z} = L/2$ and
    (b) correlation ratios.
    Here, $\eta$ and $\nu$ are critical exponents of the 3D XY model.
    $T_{c}/J \simeq 2.2021$ is obtained from the crossings of 
    dimensionless scaling parameters.
    \label{fig:DFP-FSS}
  }
\end{figure}

We first present the results for very small $\left|\lambda\right|$ with 
$\left|\lambda\right| L^{y_{\lambda}}\lesssim 1$, where we 
observe an apparently continuous transition controlled 
by the DFP. This is naturally expected from the scaling argument 
given above and  basically the same behavior that was 
observed in the frustrated model previously studied in Ref.~\onlinecite{kamiya2009finite}. In 
Fig.~\ref{fig:DFP-FSS}(a), we present the FSS plots
of $G^{m}_{ij}$ and $G^{\sigma}_{ij}$ at the largest distance in a given system where
$r_{ij,x} = r_{ij,y} = r_{ij,z} = L/2$ ($\lambda = -0.05$, $L \le 64$). These plots are based on 
the following FSS forms at the DFP \cite{kamiya2009finite}:
$
G_{ij}^{m}\left(T, L, r_{ij}\right) 
\sim 
L^{-(\eta+1)}
f_{m}(L^{1/\nu}(T - T_{c}), r_{ij}/L)
$
and
$
G_{ij}^{\sigma}\left(T, L, r_{ij}\right) 
\sim 
L^{-2(\eta+1)}
f_{\sigma}(L^{1/\nu}(T - T_{c}), r_{ij}/L)
$
with $\eta = 0.0380(4)$ and $\nu = 0.68155(27)$ being the critical
exponents of the 3D XY model \cite{campostrini2001critical}. 
Using the exponent of the DFP, we can also produce reasonable FSS 
plots for the correlation ratios $g_{m}$ and $g_{\sigma}$ 
\cite{tomita2002finite}, defined by ratios of the corresponding 
correlation functions at two different distances
$r_{ij,x} = r_{ij,y} = r_{ij,z} = L/2, L/4$ 
[see Fig. \ref{fig:DFP-FSS} (b)].

\begin{figure}
  \centering
  \includegraphics[width=8.4cm]{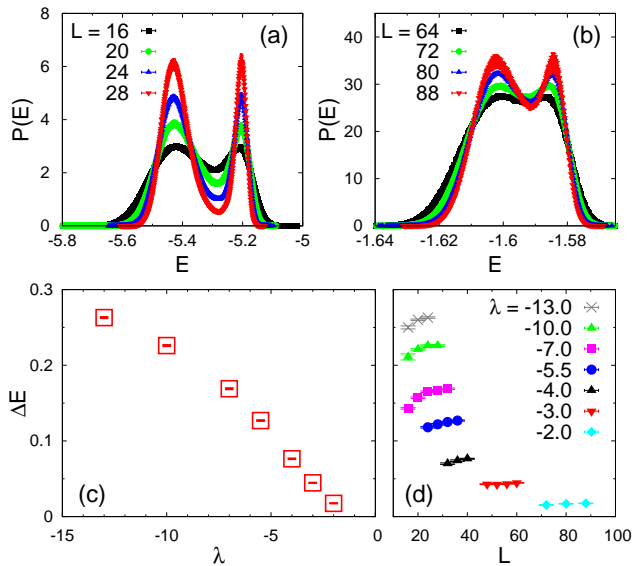}   %HERE (uncomment here for reprint)
  \caption{
    (Color online)
    Bimodal energy distribution at $T \simeq T_{c}$ for 
    (a) $\lambda = -10$ 
    ($\left|\lambda\right|L^{y_{\lambda}} \approx 43$ for $L = 16$) and 
    (b) $\lambda = -2$
    ($\left|\lambda\right|L^{y_{\lambda}} \approx 18$ for $L = 64$).
    Most error-bars are smaller than the symbol-sizes.
    (c) Peak-to-peak distance of the distribution corresponding to 
    the latent heat for the largest $L$ for each $\lambda$.
    (d) System-size dependence of the peak-to-peak distance.
    \label{fig:first-order-transition}
  }
\end{figure}

However, since the scaling argument shows that the DFP is unstable, 
we conclude that these FSS plots simply describe  
the ``pseudo-scaling'' behavior, i.e., as long as $\left|\lambda\right|$ 
is finite, significant deviations should eventually appear in 
large enough lattices. In other words, we cannot 
conclude that the transition is of second order because weak first-order
transitions can become practically indistinguishable from continuous transitions 
in the usual FSS analysis for small $L$. Indeed, for relatively large $\left|\lambda\right|$, 
we find obvious deviations from the DFP. As shown in Figs.~\ref{fig:first-order-transition}(a) and (b), the energy 
distributions near the transition show a bimodal structure with 
increasing depth for larger system sizes. This is clear evidence for a first-order
transition. The peak-to-peak distance gives an estimate of the 
latent heat $\Delta E\left(\lambda\right)$. As expected, the first-order nature 
becomes weaker for smaller $\left|\lambda\right|$  [see Fig.~\ref{fig:first-order-transition}(c)].  

\subsection{Monte Carlo renormalization group analysis}
Given our results for small and large values of $\left|\lambda\right|$, it is natural to ask if
there is a multicritical point where the first-order transition line
terminates. The dependence of $\Delta E\left(\lambda\right)$ on small values of
$\left|\lambda\right|$ does not provide an efficient way of answering this question
because larger lattices are required to detect smaller values of
$\Delta E$. In what follows, we explain our method to 
investigate the correct asymptotic behavior for very small $\left|\lambda\right|$. 
Our approach is a sort of MC renormalization group 
analysis \cite{hukushima2000random, itakura2003monte}. A similar technique was applied, for instance, to the random-bond 
Ising model by Hukushima and it was found that the method is very useful to obtain 
qualitative structure of the phase diagram \cite{hukushima2000random}.

We consider several dimensionless scaling parameters $R\left(\lambda\right)$
(such as $g_{m}$ and $g_{\sigma}$ defined above) and introduce their
$L$-dependent estimators $R\left(\lambda, L\right)$ as the crossings of
temperature-dependent curves of the parameters for two successive system-sizes $L$ and $2L$. 
Because the $L\to\infty$ limit, $R\left(\lambda\right)$, is expected to be scale-invariant and 
universal for a second-order transition, 
$R\left(\lambda, L\right)$ must converge to such a 
universal value if the transition is continuous. Consequently, if 
a multicritical point exists, the ``flow'' structure of 
$R\left(\lambda, L\right)$ should have a separatrix and a stable fixed
point. Here, the term ``flow'' refers to the evolution
of $R\left(\lambda, L\right)$ with increasing $L$.

\begin{figure}
  \centering
  \includegraphics[width=8.6cm]{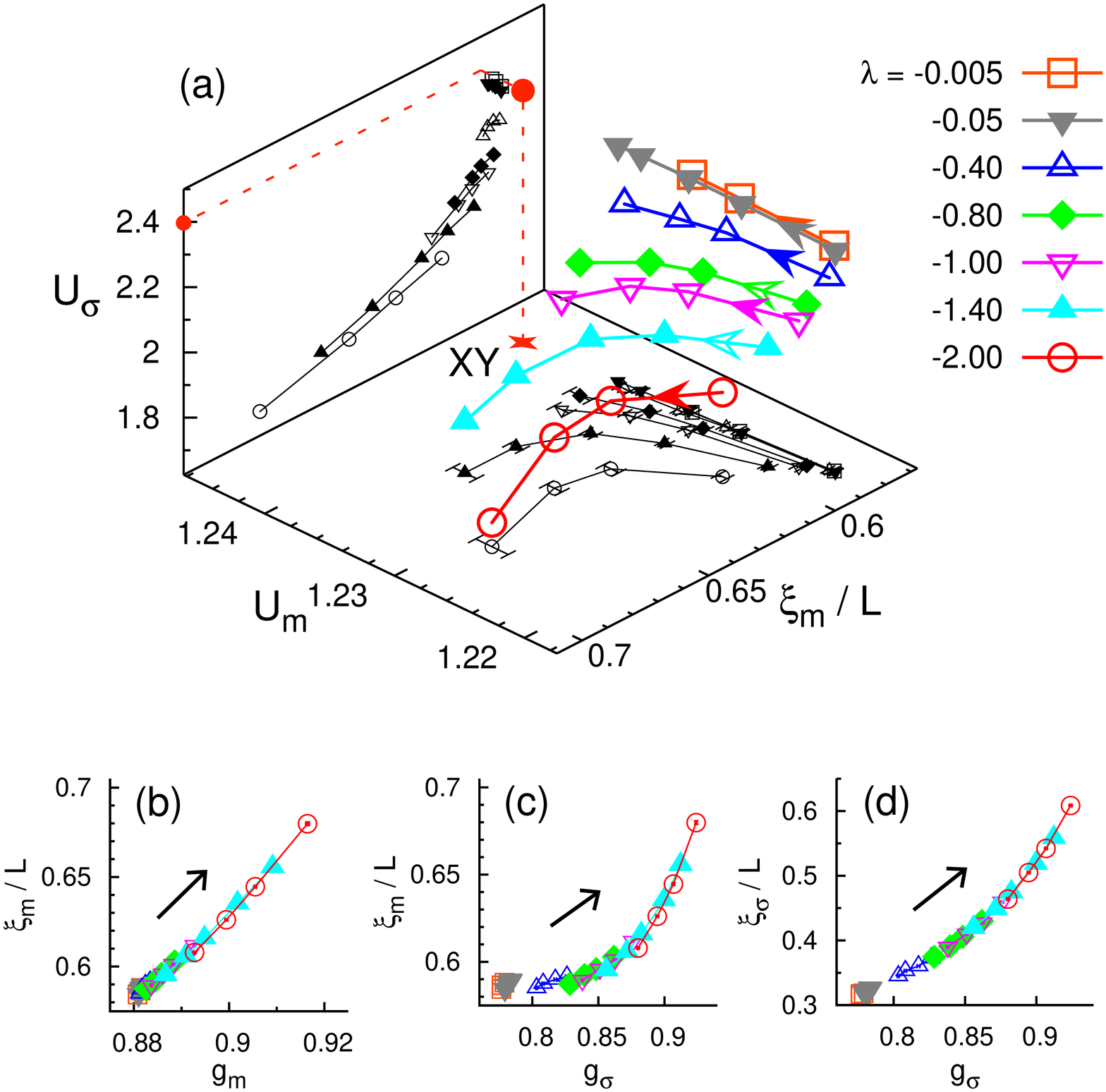}   %HERE (uncomment here for reprint)
  \caption{
    (Color online)
    (a) The flow diagram projected onto the $(U_{m}, \xi_{m}/L, U_{\sigma})$
    space. Two-dimensional projections are also shown. Error bars for $U_{m}$
    are shown on the bottom plane and those for the other parameters are 
    smaller than the symbol size (not shown). System-sizes corresponding to 
    the data points in each flow are $L = 8, 12, 16, 24$ 
    (not for $\lambda = -0.005$) and $32$ (only for $\lambda = -0.05, -1.4$), in 
    order specified by arrows attached to the flow lines. The DFP projected on the 
    $(U_{m}, {\xi}_{m}/L)$ plane is denoted by ``XY.'' 
    $U_{\sigma} \approx 2.40(1)$ at the DFP is estimated by extrapolating the 
    $\lambda = -0.005$ flow and it is shown by a small filled circle on the 
    $U_{\sigma}$ axis.
    (b)--(d) The same flow diagrams projected on the other subspaces. The 
    arrows show the overall direction of the flows.
    \label{fig:flowgrams}
  }
\end{figure}

In addition to $g_{m}$ and $g_{\sigma}$, we use as $R\left(\lambda, L\right)$
the Binder parameters defined by
$
U_{m} = 
\langle|\bm{\mathrm{m}}|^4\rangle/
\langle|\bm{\mathrm{m}}|^2\rangle^2
$
and
$
U_{\sigma} 
= \langle\sigma^4\rangle
/\langle\sigma^2\rangle^2
$,
and the second-moment correlation-lengths \cite{cooper1982solving}
divided by the system-size $\xi_{m}/L$ and $\xi_{\sigma}/L$. Hence, 
the entire parameter space is six-dimensional in our treatment. The 
obtained flow diagrams are shown in Fig.~\ref{fig:flowgrams}. 
As can be seen in Figs.~\ref{fig:flowgrams}(b--d), we 
find that in the 4D subspace
$(g_{m}, \, \xi_{m}/L, \, g_{\sigma}, \,\xi_{\sigma}/L)$
trajectories of the projected flows collapse on an 
approximately single, monotonous curve.
Therefore, it turns out to be sufficient to treat the projected flow in the
subspace spanned by one of the above four (we choose $\xi_{m}/L$)
and the other two parameters not included here, 
namely $U_{m}$ and $U_{\sigma}$. 

The flow projected onto this 
$
\left( U_{m}, \xi_{m}/L, U_{\sigma} \right)
$ 
subspace is shown in Fig.~\ref{fig:flowgrams}(a). The DFP is associated with the flows 
for $\lambda = -0.005$ or $-0.05$, because,
as is implied by the data-collapse in the FSS plots shown in Fig.~\ref{fig:DFP-FSS},
with such small $\lvert\lambda\rvert$ the effect of the biquadratic perturbation is still
negligible in the length-scale under consideration. 
The known estimates for the 3D XY universality class are 
$U_{m} = 1.2430(5)$ and $\xi_{\sigma}/L = 0.5925(2)$ 
\cite{campostrini2001critical}.
We show the point corresponding to these values on the
$(U_{m}, {\xi}_{m}/L)$ plane in Fig.~\ref{fig:flowgrams}(a).
(Estimates for the other less common parameters are not available in the
literature as far as we know.) The above observation is in good 
agreement with these estimates.

For larger values of $\left|\lambda\right|$ with $\left|\lambda\right| L^{y_{\lambda}}\gtrsim 1$, 
the flow clearly deviates from the trajectory dominated by the DFP.
This is a clear sign of the expected crossover. The crossover is already evident for 
$\lambda = -0.4$  
($\left|\lambda\right|L^{y_{\lambda}} \approx 3.1$ for $L = 48$). 
As $\left|\lambda\right|$ increases, the flow keeps evolving away from the DFP 
without a stable fixed point or a separatrix. Note that we have already shown clear evidence of a
first-order  transition for $\lambda = -2$ 
[Fig.~\ref{fig:first-order-transition}(b)]. This indicates  
that the observed crossover eventually leads to the first-order transition.

While the numerical evidence in finite systems is always insufficient for
very small $\left|\lambda\right|$, we take the numerical result presented above as a strong
evidence for the first-order character of arbitrary small $\left|\lambda\right|$.
This conclusion is also supported by the epsilon expansion analysis of 
$H_{\text{LGW}}$ around the DFP \cite{aharony1975critical}: the result
obtained by expanding the Hamiltonian to $O(\epsilon)$ is most
naturally explained as the lack of a separatrix fixed point, suggesting a fluctuation-induced first-order transition.
%While the  most natural interpretation of the flow 
%diagram is absence of a second-order transition, the numerical evidence
%on finite size systems will always be insufficient for arbitrarily small $\left|\lambda\right|$. 
%However, the absence of a continuous transition provides the 
%simplest explanation for the epsilon expansion analysis of 
%$H_{\text{LGW}}$ around the DFP \cite{aharony1975critical}: the 
%lack of a stable fixed point after expanding the Hamiltonian to order $O(\epsilon)$ 
%is usually interpreted as a fluctuation-induced first-order transition. Therefore, we conclude 
%that the crossover from the DFP finally leads to a first-order transition for any finite value of $\lambda$.

\section{Summary}
To summarize, we have established the crossover behavior from 
the 3D XY DFP for an effective model that is relevant for several frustrated 
magnets  near their thermodynamic phase transitions. Such crossover 
results in a weakly first-order phase transition. Our calculation also 
shows that it will be very difficult to observe such a first order 
transition with standard experimental methods as long as the frustrated 
inter-layer coupling is small in comparison with the intra-layer exchange. 
This is indeed the case of BaCuSi$_2$O$_6$ \cite{Ruegg2007Multiple} as
discussed in Ref.~\onlinecite{kamiya2009finite}. In 
other words, although the correct asymptotic behavior is the 
first-order transition, the thermodynamic behavior will be 
dominated by the 3D XY DFP in a broad region near the transition.
The true discontinuous nature of the transition can be observed in
a very narrow region near the transition point that could easily be
beyond the experimental precision in most cases. Nevertheless, the 
first order transition should be observable for frustrated magnets with 
$\left|\lambda\right|$ of order one.
In such cases, the 3D XY-like behavior 
beyond a certain distance from the transition point
will be finally interrupted by the fluctuation-induced first-order transition.

A value of $\left|\lambda\right|$ of order one is indeed realized in the
frustrated spin model that has been proposed for describing the iron based
superconductors LaFeAs(O$_{1-x}$F$_x$) \cite{xu2008ising,fang2008theroy}. According to our 
result, such a model should exhibit a single weakly first-order transition to
the broken O(2)$\times$Z$_{2}$ phase in presence of a strong magnetic field
(the field is required to induce effective O(2) magnetic moments). The stacked
triangular antiferromagnetic  compounds \cite{kawamura1998} are other physical
realizations of the effective model considered here
[Eq.(\ref{eq:lattice-hamiltonian})]. Similarly, we predict a single weakly
first-order phase transition to take place in these systems in presence of a
strong magnetic field, which is in agreement with recent investigations 
\cite{itakura2003monte, ngo2008stacked}.

\begin{acknowledgments}
We would like to thank M. Oshikawa and Y. Tomita for illuminating suggestions.
The computation in the present work is executed on computers at the
Supercomputer Center, Institute for Solid State Physics, University of Tokyo,
and T2K Open Supercomputer, University of Tokyo.
The project is supported by 
the MEXT Global COE Program ``the Physical Science Frontier,'' 
the MEXT Grand-in-Aid for Scientific Research (B) (22340111), 
the MEXT Grand-in-Aid for Scientific Research on Priority Areas
``Novel States of Matter Induced by Frustration'' (19052004), 
and by the Next Generation Supercomputing Project, Nanoscience Program, MEXT, Japan.
\end{acknowledgments}

\bibliography{ref}

\begin{thebibliography}{25}
\expandafter\ifx\csname natexlab\endcsname\relax\def\natexlab#1{#1}\fi
\expandafter\ifx\csname bibnamefont\endcsname\relax
  \def\bibnamefont#1{#1}\fi
\expandafter\ifx\csname bibfnamefont\endcsname\relax
  \def\bibfnamefont#1{#1}\fi
\expandafter\ifx\csname citenamefont\endcsname\relax
  \def\citenamefont#1{#1}\fi
\expandafter\ifx\csname url\endcsname\relax
  \def\url#1{\texttt{#1}}\fi
\expandafter\ifx\csname urlprefix\endcsname\relax\def\urlprefix{URL }\fi
\providecommand{\bibinfo}[2]{#2}
\providecommand{\eprint}[2][]{\url{#2}}

\bibitem[{\citenamefont{Kageyama et~al.}(1999)\citenamefont{Kageyama,
  Yoshimura, Stern, Mushnikov, Onizuka, Kato, Kosuge, Slichter, Goto, and
  Ueda}}]{kageyama1999exact}
\bibinfo{author}{\bibfnamefont{H.}~\bibnamefont{Kageyama}},
  \bibinfo{author}{\bibfnamefont{K.}~\bibnamefont{Yoshimura}},
  \bibinfo{author}{\bibfnamefont{R.}~\bibnamefont{Stern}},
  \bibinfo{author}{\bibfnamefont{N.~V.} \bibnamefont{Mushnikov}},
  \bibinfo{author}{\bibfnamefont{K.}~\bibnamefont{Onizuka}},
  \bibinfo{author}{\bibfnamefont{M.}~\bibnamefont{Kato}},
  \bibinfo{author}{\bibfnamefont{K.}~\bibnamefont{Kosuge}},
  \bibinfo{author}{\bibfnamefont{C.~P.} \bibnamefont{Slichter}},
  \bibinfo{author}{\bibfnamefont{T.}~\bibnamefont{Goto}}, \bibnamefont{and}
  \bibinfo{author}{\bibfnamefont{Y.}~\bibnamefont{Ueda}},
  \bibinfo{journal}{Phys. Rev. Lett.} \textbf{\bibinfo{volume}{82}},
  \bibinfo{pages}{3168} (\bibinfo{year}{1999}).

\bibitem[{\citenamefont{Kodama et~al.}(2002)\citenamefont{Kodama, Takigawa,
  Horvatic, Berthier, Kageyama, Ueda, Miyahara, Becca, and
  Mila}}]{kodama2002magnetic}
\bibinfo{author}{\bibfnamefont{K.}~\bibnamefont{Kodama}},
  \bibinfo{author}{\bibfnamefont{M.}~\bibnamefont{Takigawa}},
  \bibinfo{author}{\bibfnamefont{M.}~\bibnamefont{Horvatic}},
  \bibinfo{author}{\bibfnamefont{C.}~\bibnamefont{Berthier}},
  \bibinfo{author}{\bibfnamefont{H.}~\bibnamefont{Kageyama}},
  \bibinfo{author}{\bibfnamefont{Y.}~\bibnamefont{Ueda}},
  \bibinfo{author}{\bibfnamefont{S.}~\bibnamefont{Miyahara}},
  \bibinfo{author}{\bibfnamefont{F.}~\bibnamefont{Becca}}, \bibnamefont{and}
  \bibinfo{author}{\bibfnamefont{F.}~\bibnamefont{Mila}},
  \bibinfo{journal}{Science} \textbf{\bibinfo{volume}{298}},
  \bibinfo{pages}{395} (\bibinfo{year}{2002}).

\bibitem[{\citenamefont{Sebastian et~al.}(2008)\citenamefont{Sebastian,
  Harrison, Sengupta, Batista, Francoual, Palm, Murphy, Marcano, Dabkowska, and
  Gaulin}}]{sebastian2008fractalization}
\bibinfo{author}{\bibfnamefont{S.~E.} \bibnamefont{Sebastian}},
  \bibinfo{author}{\bibfnamefont{N.}~\bibnamefont{Harrison}},
  \bibinfo{author}{\bibfnamefont{P.}~\bibnamefont{Sengupta}},
  \bibinfo{author}{\bibfnamefont{C.~D.} \bibnamefont{Batista}},
  \bibinfo{author}{\bibfnamefont{S.}~\bibnamefont{Francoual}},
  \bibinfo{author}{\bibfnamefont{E.}~\bibnamefont{Palm}},
  \bibinfo{author}{\bibfnamefont{T.}~\bibnamefont{Murphy}},
  \bibinfo{author}{\bibfnamefont{N.}~\bibnamefont{Marcano}},
  \bibinfo{author}{\bibfnamefont{H.~A.} \bibnamefont{Dabkowska}},
  \bibnamefont{and} \bibinfo{author}{\bibfnamefont{B.~D.}
  \bibnamefont{Gaulin}}, \bibinfo{journal}{Proc. Natl. Acad. Sci. USA}
  \textbf{\bibinfo{volume}{105}}, \bibinfo{pages}{20157}
  (\bibinfo{year}{2008}).

\bibitem[{\citenamefont{Bulaevskii et~al.}(2008)\citenamefont{Bulaevskii,
  Batista, Mostovoy, and Khomskii}}]{bulaevskii2008electronic}
\bibinfo{author}{\bibfnamefont{L.~N.} \bibnamefont{Bulaevskii}},
  \bibinfo{author}{\bibfnamefont{C.~D.} \bibnamefont{Batista}},
  \bibinfo{author}{\bibfnamefont{M.~V.} \bibnamefont{Mostovoy}},
  \bibnamefont{and} \bibinfo{author}{\bibfnamefont{D.~I.}
  \bibnamefont{Khomskii}}, \bibinfo{journal}{Phys. Rev. B}
  \textbf{\bibinfo{volume}{78}}, \bibinfo{pages}{024402}
  (\bibinfo{year}{2008}).

\bibitem[{\citenamefont{Sebastian et~al.}(2006)\citenamefont{Sebastian,
  Harrison, Batista, Balicas, Jaime, Sharma, Kawashima, and
  Fisher}}]{sebastian2006dimensional}
\bibinfo{author}{\bibfnamefont{S.~E.} \bibnamefont{Sebastian}},
  \bibinfo{author}{\bibfnamefont{N.}~\bibnamefont{Harrison}},
  \bibinfo{author}{\bibfnamefont{C.~D.} \bibnamefont{Batista}},
  \bibinfo{author}{\bibfnamefont{L.}~\bibnamefont{Balicas}},
  \bibinfo{author}{\bibfnamefont{M.}~\bibnamefont{Jaime}},
  \bibinfo{author}{\bibfnamefont{P.~A.} \bibnamefont{Sharma}},
  \bibinfo{author}{\bibfnamefont{N.}~\bibnamefont{Kawashima}},
  \bibnamefont{and} \bibinfo{author}{\bibfnamefont{I.~R.}
  \bibnamefont{Fisher}}, \bibinfo{journal}{Nature}
  \textbf{\bibinfo{volume}{441}}, \bibinfo{pages}{617} (\bibinfo{year}{2006}).

\bibitem[{\citenamefont{Delamotte et~al.}(2008)\citenamefont{Delamotte,
  Holovatch, Ivaneyko, Mouhanna, and Tissier}}]{delamotte2008fixed}
\bibinfo{author}{\bibfnamefont{B.}~\bibnamefont{Delamotte}},
  \bibinfo{author}{\bibfnamefont{Y.}~\bibnamefont{Holovatch}},
  \bibinfo{author}{\bibfnamefont{D.}~\bibnamefont{Ivaneyko}},
  \bibinfo{author}{\bibfnamefont{D.}~\bibnamefont{Mouhanna}}, \bibnamefont{and}
  \bibinfo{author}{\bibfnamefont{M.}~\bibnamefont{Tissier}},
  \bibinfo{journal}{J. Stat. Mech.: Theory Exp.}
  (\bibinfo{year}{2008}) \bibinfo{pages}{P03014}.

\bibitem[{\citenamefont{Hukushima}(2000)}]{hukushima2000random}
\bibinfo{author}{\bibfnamefont{K.}~\bibnamefont{Hukushima}},
  \bibinfo{journal}{J. Phys. Soc. Jpn.} \textbf{\bibinfo{volume}{69}},
  \bibinfo{pages}{631} (\bibinfo{year}{2000}).

\bibitem[{\citenamefont{Itakura}(2003)}]{itakura2003monte}
\bibinfo{author}{\bibfnamefont{M.}~\bibnamefont{Itakura}}, \bibinfo{journal}{J.
  Phys. Soc. Jpn.} \textbf{\bibinfo{volume}{72}}, \bibinfo{pages}{74}
  (\bibinfo{year}{2003}).

\bibitem[{\citenamefont{Kamiya et~al.}(2009)\citenamefont{Kamiya, Kawashima,
  and Batista}}]{kamiya2009finite}
\bibinfo{author}{\bibfnamefont{Y.}~\bibnamefont{Kamiya}},
  \bibinfo{author}{\bibfnamefont{N.}~\bibnamefont{Kawashima}},
  \bibnamefont{and} \bibinfo{author}{\bibfnamefont{C.~D.}
  \bibnamefont{Batista}}, \bibinfo{journal}{J. Phys. Soc. Jpn.}
  \textbf{\bibinfo{volume}{78}}, \bibinfo{pages}{094008}
  (\bibinfo{year}{2009}).

\bibitem[{\citenamefont{Henley}(1989)}]{henley1989ordering}
\bibinfo{author}{\bibfnamefont{C.~L.} \bibnamefont{Henley}},
  \bibinfo{journal}{Phys. Rev. Lett.} \textbf{\bibinfo{volume}{62}},
  \bibinfo{pages}{2056} (\bibinfo{year}{1989}).

\bibitem[{\citenamefont{Chandra et~al.}(1990)\citenamefont{Chandra, Coleman,
  and Larkin}}]{chandra1990ising}
\bibinfo{author}{\bibfnamefont{P.}~\bibnamefont{Chandra}},
  \bibinfo{author}{\bibfnamefont{P.}~\bibnamefont{Coleman}}, \bibnamefont{and}
  \bibinfo{author}{\bibfnamefont{A.~I.} \bibnamefont{Larkin}},
  \bibinfo{journal}{Phys. Rev. Lett.} \textbf{\bibinfo{volume}{64}},
  \bibinfo{pages}{88} (\bibinfo{year}{1990}).

\bibitem[{\citenamefont{Batista et~al.}(2007)\citenamefont{Batista, Schmalian,
  Kawashima, Sengupta, Sebastian, Harrison, Jaime, and
  Fisher}}]{batista2007geometric}
\bibinfo{author}{\bibfnamefont{C.~D.} \bibnamefont{Batista}},
  \bibinfo{author}{\bibfnamefont{J.}~\bibnamefont{Schmalian}},
  \bibinfo{author}{\bibfnamefont{N.}~\bibnamefont{Kawashima}},
  \bibinfo{author}{\bibfnamefont{P.}~\bibnamefont{Sengupta}},
  \bibinfo{author}{\bibfnamefont{S.~E.} \bibnamefont{Sebastian}},
  \bibinfo{author}{\bibfnamefont{N.}~\bibnamefont{Harrison}},
  \bibinfo{author}{\bibfnamefont{M.}~\bibnamefont{Jaime}}, \bibnamefont{and}
  \bibinfo{author}{\bibfnamefont{I.~R.} \bibnamefont{Fisher}},
  \bibinfo{journal}{Phys. Rev. Lett.} \textbf{\bibinfo{volume}{98}},
  \bibinfo{pages}{257201} (\bibinfo{year}{2007}).

\bibitem[{\citenamefont{R{\"o}sch and Vojta}(2007)}]{rosch2007quantum}
\bibinfo{author}{\bibfnamefont{O.}~\bibnamefont{R{\"o}sch}} \bibnamefont{and}
  \bibinfo{author}{\bibfnamefont{M.}~\bibnamefont{Vojta}},
  \bibinfo{journal}{Phys. Rev. B} \textbf{\bibinfo{volume}{76}},
  \bibinfo{pages}{180401(R)} (\bibinfo{year}{2007}).

\bibitem[{\citenamefont{Schmalian and Batista}(2008)}]{schmalian2008emergent}
\bibinfo{author}{\bibfnamefont{J.}~\bibnamefont{Schmalian}} \bibnamefont{and}
  \bibinfo{author}{\bibfnamefont{C.~D.} \bibnamefont{Batista}},
  \bibinfo{journal}{Phys. Rev. B} \textbf{\bibinfo{volume}{77}},
  \bibinfo{pages}{094406} (\bibinfo{year}{2008}).

\bibitem[{\citenamefont{Hasenbusch et~al.}(2005)\citenamefont{Hasenbusch,
  Pelissetto, and Vicari}}]{hasenbusch2005multicritical}
\bibinfo{author}{\bibfnamefont{M.}~\bibnamefont{Hasenbusch}},
  \bibinfo{author}{\bibfnamefont{A.}~\bibnamefont{Pelissetto}},
  \bibnamefont{and} \bibinfo{author}{\bibfnamefont{E.}~\bibnamefont{Vicari}},
  \bibinfo{journal}{J. Stat. Mech.: Theory Exp.} (\bibinfo{year}{2005})
  \bibinfo{pages}{P12002}.

\bibitem[{\citenamefont{Aharony}(1975)}]{aharony1975critical}
\bibinfo{author}{\bibfnamefont{A.}~\bibnamefont{Aharony}},
  \bibinfo{journal}{Phys. Rev. B} \textbf{\bibinfo{volume}{12}},
  \bibinfo{pages}{1038} (\bibinfo{year}{1975}).

\bibitem[{\citenamefont{Wolff}(1989)}]{wolff1989collective}
\bibinfo{author}{\bibfnamefont{U.}~\bibnamefont{Wolff}},
  \bibinfo{journal}{Phys. Rev. Lett.} \textbf{\bibinfo{volume}{62}},
  \bibinfo{pages}{361} (\bibinfo{year}{1989}).

\bibitem[{\citenamefont{Campostrini et~al.}(2001)\citenamefont{Campostrini,
  Hasenbusch, Pelissetto, Rossi, and Vicari}}]{campostrini2001critical}
\bibinfo{author}{\bibfnamefont{M.}~\bibnamefont{Campostrini}},
  \bibinfo{author}{\bibfnamefont{M.}~\bibnamefont{Hasenbusch}},
  \bibinfo{author}{\bibfnamefont{A.}~\bibnamefont{Pelissetto}},
  \bibinfo{author}{\bibfnamefont{P.}~\bibnamefont{Rossi}}, \bibnamefont{and}
  \bibinfo{author}{\bibfnamefont{E.}~\bibnamefont{Vicari}},
  \bibinfo{journal}{Phys. Rev. B} \textbf{\bibinfo{volume}{63}},
  \bibinfo{pages}{214503} (\bibinfo{year}{2001}).

\bibitem[{\citenamefont{Tomita and Okabe}(2002)}]{tomita2002finite}
\bibinfo{author}{\bibfnamefont{Y.}~\bibnamefont{Tomita}} \bibnamefont{and}
  \bibinfo{author}{\bibfnamefont{Y.}~\bibnamefont{Okabe}},
  \bibinfo{journal}{Phys. Rev. B} \textbf{\bibinfo{volume}{66}},
  \bibinfo{pages}{180401(R)} (\bibinfo{year}{2002}).

\bibitem[{\citenamefont{Cooper et~al.}(1982)\citenamefont{Cooper, Freedman, and
  Preston}}]{cooper1982solving}
\bibinfo{author}{\bibfnamefont{F.}~\bibnamefont{Cooper}},
  \bibinfo{author}{\bibfnamefont{B.}~\bibnamefont{Freedman}}, \bibnamefont{and}
  \bibinfo{author}{\bibfnamefont{D.}~\bibnamefont{Preston}},
  \bibinfo{journal}{Nucl. Phys. B} \textbf{\bibinfo{volume}{210}},
  \bibinfo{pages}{210} (\bibinfo{year}{1982}).

\bibitem[{\citenamefont{R\"uegg et~al.}(2007)\citenamefont{R\"uegg, McMorrow,
  Normand, R\o{}nnow, Sebastian, Fisher, Batista, Gvasaliya, Niedermayer, and
  Stahn}}]{Ruegg2007Multiple}
\bibinfo{author}{\bibfnamefont{C.}~\bibnamefont{R\"uegg}},
  \bibinfo{author}{\bibfnamefont{D.~F.} \bibnamefont{McMorrow}},
  \bibinfo{author}{\bibfnamefont{B.}~\bibnamefont{Normand}},
  \bibinfo{author}{\bibfnamefont{H.~M.} \bibnamefont{R\o{}nnow}},
  \bibinfo{author}{\bibfnamefont{S.~E.} \bibnamefont{Sebastian}},
  \bibinfo{author}{\bibfnamefont{I.~R.} \bibnamefont{Fisher}},
  \bibinfo{author}{\bibfnamefont{C.~D.} \bibnamefont{Batista}},
  \bibinfo{author}{\bibfnamefont{S.~N.} \bibnamefont{Gvasaliya}},
  \bibinfo{author}{\bibfnamefont{C.}~\bibnamefont{Niedermayer}},
  \bibnamefont{and} \bibinfo{author}{\bibfnamefont{J.}~\bibnamefont{Stahn}},
  \bibinfo{journal}{Phys. Rev. Lett.} \textbf{\bibinfo{volume}{98}},
  \bibinfo{pages}{017202} (\bibinfo{year}{2007}).

\bibitem[{\citenamefont{Xu et~al.}(2008)\citenamefont{Xu, M\"uller, and
  Sachdev}}]{xu2008ising}
\bibinfo{author}{\bibfnamefont{C.}~\bibnamefont{Xu}},
  \bibinfo{author}{\bibfnamefont{M.}~\bibnamefont{M\"uller}}, \bibnamefont{and}
  \bibinfo{author}{\bibfnamefont{S.}~\bibnamefont{Sachdev}},
  \bibinfo{journal}{Phys. Rev. B} \textbf{\bibinfo{volume}{78}},
  \bibinfo{pages}{020501(R)} (\bibinfo{year}{2008}).

\bibitem[{\citenamefont{Fang et~al.}(2008)\citenamefont{Fang, Yao, Tsai, Hu,
  and Kivelson}}]{fang2008theroy}
\bibinfo{author}{\bibfnamefont{C.}~\bibnamefont{Fang}},
  \bibinfo{author}{\bibfnamefont{H.}~\bibnamefont{Yao}},
  \bibinfo{author}{\bibfnamefont{W.-F.} \bibnamefont{Tsai}},
  \bibinfo{author}{\bibfnamefont{J.~P.}~\bibnamefont{Hu}}, \bibnamefont{and}
  \bibinfo{author}{\bibfnamefont{S.~A.} \bibnamefont{Kivelson}},
  \bibinfo{journal}{Phys. Rev. B} \textbf{\bibinfo{volume}{77}},
  \bibinfo{pages}{224509} (\bibinfo{year}{2008}).

\bibitem[{\citenamefont{Kawamura}(1998)}]{kawamura1998}
\bibinfo{author}{\bibfnamefont{H.}~\bibnamefont{Kawamura}},
  \bibinfo{journal}{J. Phys.: Condens. Matter} \textbf{\bibinfo{volume}{10}},
  \bibinfo{pages}{4707} (\bibinfo{year}{1998}).

\bibitem[{\citenamefont{Ngo and Diep}(2008)}]{ngo2008stacked}
\bibinfo{author}{\bibfnamefont{V.}~\bibnamefont{Ngo}} \bibnamefont{and}
  \bibinfo{author}{\bibfnamefont{H.}~\bibnamefont{Diep}}, \bibinfo{journal}{J.
  Appl. Phys.} \textbf{\bibinfo{volume}{103}}, \bibinfo{pages}{07C712}
  (\bibinfo{year}{2008}).

\end{thebibliography}

\end{document}